\documentclass[a4paper]{jpconf}
\usepackage{graphicx}
\usepackage{amssymb}
\usepackage{amsmath}
\usepackage{bm}
\usepackage{color}
\newcommand{\ua}{\uparrow}
\newcommand{\da}{\downarrow}
\newcommand{\dg}{\dagger}

\begin{document}
\title{A variational Monte Carlo study of exciton condensation}

\author{Hiroshi Watanabe$^{1}$, Kazuhiro Seki$^{2}$, and Seiji Yunoki$^{1,2,3}$}

\address{$^1$Computational Quantum Matter Research Team, RIKEN Center for Emergent Matter Science (CEMS), Wako, Saitama 351-0198, Japan}
\address{$^2$Computational Condensed Matter Physics Laboratory, RIKEN, Saitama 351-0198, Japan}
\address{$^3$Computational Materials Science Research Team, RIKEN Advanced Institute for 
Computational Science (AICS), Kobe, Hyogo 650-0047, Japan}

\ead{h-watanabe@riken.jp}

\begin{abstract}
Exciton condensation in a two-band Hubbard model on a square lattice is studied with variational Monte Carlo method.
We show that the phase transition from an excitonic insulator to a band insulator is induced by increasing the interband Coulomb interaction.
To examine the character of the exciton condensation, the exciton pair amplitudes both in $\bm{k}$-space and in real space are calculated.
Using these quantities, we discuss the BCS-BEC crossover within the excitonic insulator phase.
\end{abstract}

\section{Introduction}
Exciton condensation is one of the most interesting phenomena in strongly correlated electron systems and it has been extensively studied since 1960s~\cite{Mott,Jerome,Halperin}.
The exciton is a bound electron-hole pair mediated by the (attractive) Coulomb interaction between them.
When the binding energy of an exciton exceeds the energy gap between the electron and hole bands, the system has an instability towards condensation of excitons.
Despite the extensive studies for half a century, the example of exciton condensation which is generally accepted is quite limited.
One of the reasons is that the exciton is a neutral quasiparticle and difficult to be detected in experiments.
Therefore, any conclusive evidence for exciton condensation in real materials is essential for further progress.

Among the limited candidates, the transition metal dichalcogenide of TiSe$_2$ is one of the most notable materials for exciton 
condensation. It shows a commensurate CDW transition at $T_c\sim$200K with small structural distortion but the Fermi surface partially remains due to the imperfect opening of the energy gap.
The origin of the CDW transition is still controversial and one of the possible mechanisms is exciton condensation~\cite{Monney,Cazzaniga}. 
The observed flat band structure just below the Fermi energy~\cite{Cercellier} is characteristic of exciton condensation, although it is not a direct evidence.
Another possible mechanism for the CDW transition is a Jahn-Teller effect which results from the electron-phonon interaction~\cite{Hughes}.
Recently, it is proposed that both mechanisms work cooperatively for the CDW state~\cite{vanWezel,Zenker}.
In addition to the CDW phase, TiSe$_2$ exhibits superconductivity with dome-shaped transition temperature induced by intercalating Cu atoms between layers~\cite{Morosan} or by applying pressure~\cite{Kusmartseva}.
The origin of the superconductivity has not been clarified yet and the relation between the CDW and the superconductivity has attracted much interests in the context of the
quantum critical point.

Most of the theoretical studies for exciton condensation are limited to an ideal case, e.g., considering models in one dimensional,  ignoring an intraband Coulomb interaction, a long-range Coulomb interaction, and a spin degrees of freedom. 
Although these models are useful to describe exciton condensation alone, it is  generally difficult to discuss the competition between other competing phases such as magnetism and superconductivity due to the lack of realistic condition.
Our aim is to establish a calculation scheme which can describe such a competition and to clarify the emerging mechanism of exciton condensation and superconductivity in low carrier density systems such as TiSe$_2$.
The variational Monte Carlo (VMC) method is one of the powerful methods for a strongly correlated electron systems and is suitable to this kind of problem.
In this paper, we study a two-band Hubbard model with intra- and interband Coulomb interactions to investigate the detailed character of  exciton condensation.
The VMC method is used for the calculation of physical quantities in the ground state.
We show that the phase transition from an excitonic insulator to a band insulator is induced by increasing the interband interaction.
The importance of the intraband interaction is also discussed.
Furthermore, the BCS-BEC crossover within the excitonic insulator is studied by calculating the exciton pair amplitudes both  in $\bm{k}$-space and in real space.

\section{Model and method}
\begin{figure}[t]
\begin{center}
\includegraphics[width=0.85\linewidth]{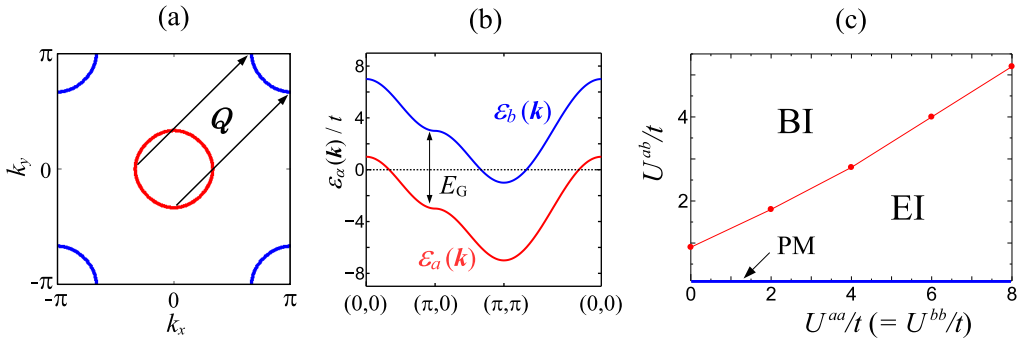}
\caption{(color online) (a) Fermi surfaces and (b) energy dispersions of the noninteracting tight-binding energy band with electron density $n=2$.
(c) Ground state phase diagram in the $U^{aa}/t-U^{ab}/t$ plane. PM, EI, and BI denote paramagnetic metal, excitonic insulator, and band insulator, respectively.
PM is stabilized only for $U^{ab}/t=0$ due to the perfect nesting condition.
\label{fig1}} 
\end{center}
\end{figure}

We consider a two-band Hubbard model on a two-dimensional square lattice defined by the following Hamiltonian,
\begin{equation}
H=\sum_{\bm{k}\sigma}\varepsilon_a(\bm{k})a^{\dg}_{\bm{k}\sigma}a_{\bm{k}\sigma}
+\sum_{\bm{k}\sigma}\varepsilon_b(\bm{k})b^{\dg}_{\bm{k}\sigma}b_{\bm{k}\sigma}
+U^{aa}\sum_in^a_{i\ua}n^a_{i\da}
+U^{bb}\sum_in^b_{i\ua}n^b_{i\da}
+U^{ab}\sum_in^a_in^b_i
\end{equation}
where $\alpha^{\dg}_{\bm{k}\sigma} (\alpha=a,b)$ denotes the creation operator of an electron with wave vector $\bm{k}$ and spin 
$\sigma\,(=\uparrow,\downarrow)$ on 
band $\alpha$. $U^{\alpha\alpha}$ is an intraband Coulomb interaction and $U^{ab}$ is an interband Coulomb interaction between bands $a$ and $b$.
$n^{\alpha}_{i\sigma}=\alpha^{\dg}_{i\sigma}\alpha_{i\sigma}$ denotes the number operator and $n^{\alpha}_i=\sum_{\sigma}n^{\alpha}_{i\sigma}$.
We set the non-interacting band energy as $\varepsilon_a(\bm{k})=2t_a(\cos k_x+\cos k_y)$ and $\varepsilon_b(\bm{k})=2t_b(\cos k_x+\cos k_y)+E_{\text{G}}$.
The Fermi surface and the energy dispersion are shown in Figs.~\ref{fig1}(a) and (b), respectively.
The electron and hole Fermi surfaces are perfectly nested with wave vector $\bm{Q}=(\pi,\pi)$ and have instability towards exciton condensation.

Here, we introduce the following Gutzwiller-Jastrow type trial wave function:
\begin{equation}
\left|\Psi \right>=P_{\mathrm{J_c}}P^{(2)}_{\mathrm{G}}\left|\Phi \right>.
\label{wf}
\end{equation}
The one-body part $\left|\Phi \right>$ is obtained by diagonalizing 
the mean-field Hamiltonian,
\begin{equation}
H_{\text{MF}}=\sum_{\bm{k}\sigma}(a^{\dagger}_{\bm{k}\sigma}, b^{\dagger}_{\bm{k}+\bm{Q}\sigma})
\begin{pmatrix}
\varepsilon_a(\bm{k}) & \tilde{\Delta}_{\bm{k}} \\
\tilde{\Delta}_{\bm{k}} &\tilde{\varepsilon}_b(\bm{k}+\bm{Q})
\end{pmatrix}
\begin{pmatrix}
a_{\bm{k}\sigma} \\
b_{\bm{k}+\bm{Q}\sigma} 
\end{pmatrix}.
\end{equation}
Since we fix $t_a=t$ as an energy unit, $\varepsilon_a(\bm{k})$ is unchanged through the VMC calculation.
On the other hand, $\tilde{t}_b$ and $\tilde{E}_{\text{G}}$ in $\tilde{\varepsilon}_b(\bm{k})$ are optimized so as to minimize the variational energy.
$\tilde{\Delta}_{\bm{k}}$ is a variational parameter for the exciton condensation and generates the spontaneous hybridization between bands $a$ and $b$ with wave vector $\bm{Q}=(\pi,\pi)$.
Although there are several types of exciton condensation depending on charge and spin degrees of freedom, 
they are energetically degenerate for the Hamiltonian without spin-dependent interactions such as Hund's coupling. 
Therefore, we assume a spin-independent variational parameter as follows,
\begin{equation}
\tilde{\Delta}_{\bm{k}}=\tilde{\Delta}
\exp\biggl[-\tilde{A}(\varepsilon_a(\bm{k})
-\tilde{\varepsilon}_b(\bm{k}+\bm{Q}))^2 \biggr].
\end{equation}
$\tilde{\Delta}$ is an amplitude of the variational parameter and $\tilde{A}$ controls the extent of exciton 
in $\bm{k}$-space.
We have found that the exciton condensation can be described without $\tilde{A}$ (i.e., $\tilde{\Delta}_{\bm{k}}=\tilde{\Delta}$)
but the introduction of $\tilde{A}$ greatly improves the variational energy and gives a better trial wave function.

The Gutzwiller operator 
\begin{equation}
P^{(2)}_{\mathrm{G}}=\prod_{i,\gamma}\left[1-(1-g_\gamma)\left|\gamma\right>\left<\gamma\right|_i\right]
\label{go}
\end{equation} 
in $\left|\Psi \right>$ is the one extended for the two-band system.
$i$ is a site index and $\gamma$ represents possible electron configurations at each site, namely, 
$\left|0\right>=\left|0\;0\right>$, $\left|1\right>=\left|0\ua\right>$, $\cdots$, $\left|15\right>=\left|\ua\da\;\ua\da\right>$.
The variational parameters $g_\gamma$'s vary from 0 to 1, which control the weight of each electron configuration.  
Here, we classify the possible 16 local electron configurations into 10 groups by the local energy 
and set the same value of $g_\gamma$'s for electron configurations with the same energy. 
The explicit grouping is shown elsewhere.

The remaining operator 
\begin{equation}\label{jastrow}
P_{\mathrm{J_c}}=\exp\left[-\sum_{i\neq j \alpha\beta}v^{\alpha\beta}_{ij}n^{\alpha}_in^{\beta}_j\right]
\end{equation}
in $\left|\Psi \right> $ is the charge Jastrow factor, which controls the long-range charge correlations.
Since the on-site Coulomb interactions $U^{aa}$, $U^{bb}$, and $U^{ab}$ affect not only the on-site but also the off-site (long-range) charge correlation,
the charge Jastrow factor is essential for describing the appropriate trial wave function~\cite{Watanabe1}.
The band dependence is fully considered, namely, $v^{aa}_{ij}$, $v^{bb}_{ij}$, and $v^{ab}_{ij}$ are independently optimized.
For the spatial dependence, we assume that $v^{\alpha\beta}_{ij}$ depends only on the distances (not on the direction), $v^{\alpha\beta}_{ij}=v^{\alpha\beta}(|\bm{r}_i-\bm{r}_j|)$, 
and consider the range of $R=|\bm{r}_i-\bm{r}_j|<L/2$ for a square lattice of $N=L\times L$. 

These variational parameters,  $\tilde{\varepsilon}_b, \tilde{\Delta}_{\bm{k}}, g_{\gamma}, {v^{\alpha\beta}_{ij}}$ are simultaneously optimized so as to minimize the variational energy.
For the optimization, we employ the stochastic reconfiguration method~\cite{Sorella} which works quite effectively for the VMC method.

\section{Results}
First, we study the stability of the exciton condensation and its competing phase by changing the intra- and interband interactions.
The system size is 28$\times$28 through the paper.
Here, we set the intraband interactions $U^{aa}=U^{bb}$ for simplicity.
Figure~\ref{fig1}(c) shows the obtained phase diagram in the $U^{aa}/t-U^{ab}/t$ plane.
The order parameter of exciton condensation is defined as $\Delta=\sum_{\bm{k},\sigma}\bigl< b^{\dagger}_{\bm{k}+\bm{Q}\sigma}a_{\bm{k}\sigma}+\text{H.c.}\bigr>$ and the finite $\Delta$ leads to a broken translational symmetry
with ordering vector $\bm{Q}=(\pi, \pi)$.
(Note that $\Delta$ is different from the variational parameter $\tilde{\Delta}_{\bm{k}}$ defined in eq. (4).)
Since the Fermi surfaces of bands $a$ and $b$ are perfectly nested as shown in Fig.~\ref{fig1}(a), an infinitesimally small $U^{ab}$ induces the exciton condensation and the system becomes fully gapped.
This phase is called excitonic insulator which has been extensively studied so far.
As for the trial wave function, we have confirmed that the charge Jastrow factor $P_{\mathrm{J_c}}$ is quite important for the stability of exiton condensation through lowering the variational energy of excitonic insulator phase.
This is because the exciton pair has nonnegligible long-range component (at least several lattice constant) even with on-site Coulomb interaction alone.
The details are discussed later in Fig.~\ref{fig3}.
When $U^{ab}$ increases further, the bands $a$ and $b$ are completely decoupled and the band insulator is realized.
In the band insulator phase, the band $a$ is fully filled ($n_{a}=2$) and the band $b$ is empty ($n_{b}=0$) to avoid the energy loss of $U^{ab}$ despite the energy loss of $U^{aa}$.
Therefore, the region of the excitonic insulator becomes larger as $U^{aa}$ increases~\cite{Zocher}.
This phase diagram is consistent with the previous variational cluster approximation (VCA) study both qualitatively and quantitatively~\cite{Kaneko}.
It is interesting to note that the excitonic insulator phase is restricted to a quite narrow region when the intraband interaction is absent ($U^{aa}=0$).
This indicates the importance of the intraband interaction to stabilize the exciton condensation.
Since most of the previous studies for exciton condensation neglect the intraband Coulomb interactions, its stability might be underestimated.

\begin{figure}[t]
\begin{center}
\includegraphics[width=0.9\linewidth]{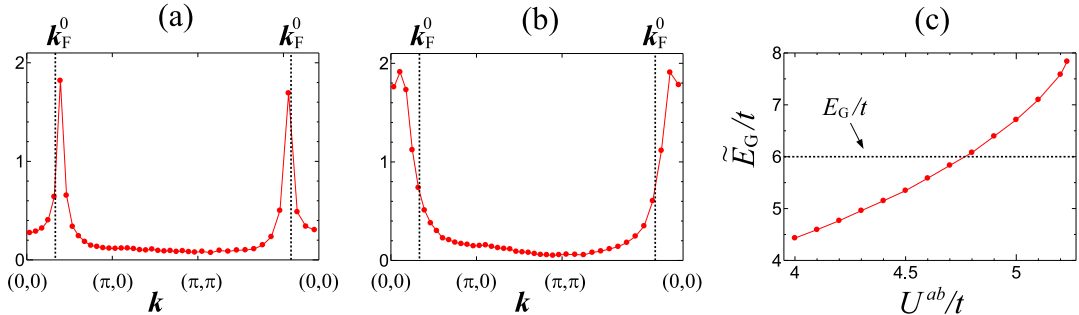}
\caption{(color online) Electron-hole pair amplitude in $\bm{k}$-space $\phi(\bm{k})$ for (a) $U^{ab}/t=4.6$ and (b) $U^{ab}/t=5.2$. 
$\bm{k}^0_{\text{F}}$ denotes the original (noninteracting) Fermi wave vector shown in Fig.~\ref{fig1}(a) with a red circle.
(c) $U^{ab}/t$ dependence of the effective energy gap $\tilde{E}_\text{G}/t$. 
Dotted line corresponds to the original value $E_\text{G}/t=6$ shown in Fig.~\ref{fig1}(b). 
In all figures, $U^{aa}/t=U^{bb}/t=8.0$.
\label{fig2}} 
\end{center}
\end{figure}

Next, we calculate the electron-hole pair amplitude in $\bm{k}$-space defined as,
\begin{equation}
\phi(\bm{k})=\sum_{\sigma}\bigl< b^{\dagger}_{\bm{k}+\bm{Q}\sigma}a_{\bm{k}\sigma}+\mathrm{H.c.}\bigr>
\end{equation}
Figure~\ref{fig2} shows the behavior of $\phi(\bm{k})$ in the first Brillouin zone.
For $U^{ab}/t=4.6$, $\phi(\bm{k})$ has a sharp peak at $\bm{k}=\tilde{\bm{k}}_{\text{F}}$, which is an effective (or renormalized) Fermi wave vector in the wave function $|\Psi\rangle$.
A sharp peak structure in $\bm{k}$-space corresponds to a widely extended electron-hole pair in real space.
The dotted line in Fig.~\ref{fig2}(a) corresponds to the original (noninteracting) Fermi wave vector $\bm{k}^0_{\text{F}}$ and we can see that
$|\tilde{\bm{k}}_{\text{F}}| > |\bm{k}^0_{\text{F}}|$.
This is because the effective energy gap $\tilde{E}_\text{G}$ in $|\Psi\rangle$ is reduced from the original value ($\tilde{E}_\text{G}/t < E_{\text{G}}/t=6$) and the Fermi surface is enlarged.
On the other hand, as shown in Fig.~\ref{fig2}(b), the peak structure is rather broad for $U^{ab}/t=5.2$ and it indicates a tightly bounded electron-hole pair in real space.
The effective Fermi wave vector is smaller than the original one, $|\tilde{\bm{k}}_{\text{F}}| < |\bm{k}^0_{\text{F}}|$, and the effective energy gap is larger
than the original one, $\tilde{E}_\text{G}/t > E_{\text{G}}/t$.
The $U^{ab}$ dependence of $\tilde{E}_\text{G}/t$ is shown in Fig.~\ref{fig2}(c) and this behavior is explained as follows. 
The character of the exciton condensation is determined by the balance between intra-($U^{aa}$) and interband ($U^{ab}$) interactions.
When the effect of $U^{aa}$ is dominant ($U^{ab}/t \lesssim$ 4.8), the effective energy gap is renormalized to a smaller value, 
$\tilde{E}_\text{G}/t < E_\text{G}/t$, to decrease $n_a$ and avoid the loss of $U^{aa}$.
As $U^{ab}$ increases for fixed $U^{aa}$, $\tilde{E}_\text{G}/t$ increases and exceeds the original value around $U^{ab}/t\sim 4.8$.
For $U^{ab}/t \gtrsim 4.8$, the effect of $U^{ab}$ is dominant and the effective energy gap is renormalized to a larger value, 
$\tilde{E}_\text{G}/t > E_\text{G}/t$, to decrease $n_b$ and avoid the loss of $U^{ab}$.
Finally, the bands $a$ and $b$ are completely decoupled and the band insulator is realized for $U^{ab}/t\gtrsim 5.24$.
Note that we cannot observe the limit of the Bose-Einstein condensation (BEC) of excitons where the effective Fermi wave vector $\tilde{\bm{k}}_{\text{F}}$ vanishes, namely,
$\tilde{E}_\text{G}/t > 8$ (no overlap between the bands $a$ and $b$) and $\Delta \neq0$.
However, the behavior of the variational parameters drastically changes across $U^{ab}/t\sim 4.8$ and we expect that there is a BCS-BEC crossover around this $U^{ab}/t$.
In the VCA study, the BCS-BEC crossover occurs around $U^{ab}/t\sim 4.0 $ and the vanishing of $\tilde{\bm{k}}_{\text{F}}$ is observed at  $U^{ab}/t =5.2$~\cite{Kaneko}. 
Although the VMC study depends on the form of trial wave functions used, it can treat large enough system sizes, as large as $N=28\times28$, 
to describe the bound electron-hole pair of exciton at least in the BEC region. 
The electron correlation is exactly treated within a 2$\times$2 cluster in the VCA, but our results suggest that the cluster size dependence 
should be carefully considered. The difference between the VMC and VCA results should be clarified to further understand the BCS-BEC crossover and it is precisely our future problem.

\begin{figure}[t]
\begin{center}
\includegraphics[width=0.8\linewidth]{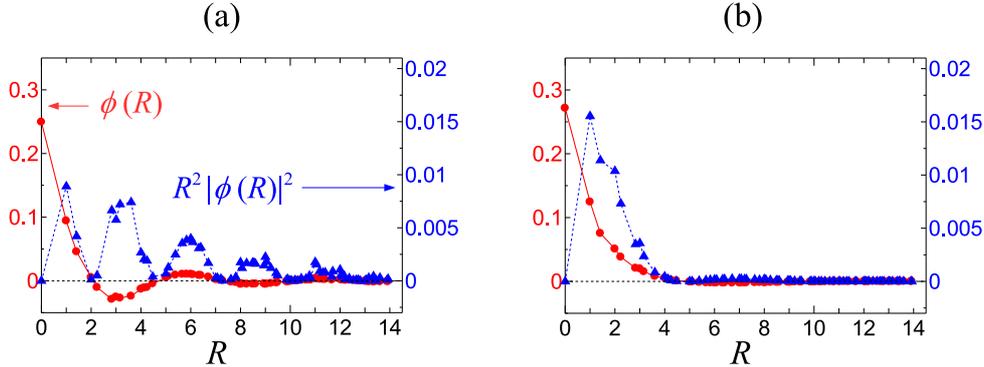}
\caption{(color online) Electron-hole pair amplitude in real space $\phi(R)$ and $R^2|\phi(R)|^2$ for (a) $U^{ab}/t=4.6$ and (b) $U^{ab}/t=5.2$ at $U^{aa}/t=U^{bb}/t=8.0$ (same with Fig.~\ref{fig2}).
\label{fig3}} 
\end{center}
\end{figure}

Next, to examine the character of exciton in real space, we calculate the electron-hole pair amplitude $\phi_{ij}$ defined as
\begin{equation}
\phi_{ij}=\sum_{\sigma}\bigl< b^{\dagger}_{i\sigma}a_{j\sigma}\bigr>+\mathrm{H.c.}
\end{equation}
Although in general $\phi_{ij}$ depends not only on the distance between $\bm{r}_i$ and $\bm{r}_j$, $R=|\bm{r}_i-\bm{r}_j|$, but also on the direction 
between $\bm{r}_i$ and $\bm{r}_j$, we have found that the direction dependence of $\phi_{ij}$ is small at least for parameters studied here.
Therefore, we assume $\phi_{ij}=\phi(R)$ in the following.
Figure~\ref{fig3} shows the behavior of $\phi(R)$ for $U^{aa}/t=U^{bb}/t=8.0$. 
For $U^{ab}/t=4.6$ (expected to be BCS type), $\phi(R)$ gradually decreases with increasing $R$ and exhibits oscillatory behavior.  
A similar oscillatory behavior can be observed in the electron-electron pair amplitude $\bigl<c^{\dg}_{i\ua}c^{\dg}_{j\da}\bigr>$ of the BCS superconductivity, which is proportional to $\sin (k_{\text{F}}R)/R$.
This is because the wave function of exciton condensation is composed of a coherent electron-hole pair and essentially identical to that of the BCS superconductivity with a coherent electron-electron pair~\cite{Jerome}.
On the other hand, for $U^{ab}/t=5.2$ (expected to be BEC type), in the vicinity of the phase boundary between the excitonic and band insulators, $\phi(R)$ decreases  more rapidly and the oscillatory behavior is hardly observed.
As $U^{ab}$ increases, the attraction between electron and hole increases and the radius of exciton becomes smaller in real space.
These behaviors are consistent with the $\bm{k}$-space picture: a widely-extended (tightly-bounded) exciton in real space corresponds to a localized (an extended) exciton in $\bm{k}$-space.

Finaly, we calculate the coherence length of electron-hole pair defined as
\begin{equation}
\xi^2=\displaystyle \sum_{R}R^2|\phi(R)|^2 \Big/
\displaystyle\sum_{R}|\phi(R)|^2. \label{coh}
\end{equation}
The coherence length is often used to distinguish the BCS region ($\xi>1$, weak coupling) from the BEC region ($\xi<1$, strong coupling).
In Fig.~\ref{fig3}, we plot the quantity $R^2|\phi(R)|^2$ appearing in Eq.(\ref{coh}).
For $U^{ab}/t=5.2$, this quantity is almost negligible 
for $R\gtrsim 4$, indicating that the system size ($N=28\times28$) is large enough to describe the bound electron-hople pair of exciton.
The estimated coherence length $\xi$ is 0.77.
On the other hand, for $U^{ab}/t=4.6$, $R^2|\phi(R)|^2$ gradually decreases with increasing $R$, showing oscillatory behavior, 
and has significant contribution for rather large $R$.
We have also found that the size dependence of the calculation becomes severe around this $U^{ab}/t$ and below. 
This indicates that there is a BCS-BEC crossover around this $U^{ab}/t$, below which the coherence length rapidly increases in the BCS 
region where even $N=28\times28$ is not large enough to estimate $\xi$.
Although it is difficult to discuss the BCS-BEC crossover with $\phi(R)$ and $\xi$ shown here, we can visualize the spatial distribution of 
exciton using these quantities, which helps for further understanding the exciton condensation.
The accurate estimation of $\xi$ and other physical quantities in real space is an important future problem.

\section{Summary and discussion}
In this paper, we have studied the two-band Hubbard model with perfectly nested electron and hole Fermi surfaces to discuss the stability of the exciton condensation with changing the Coulomb interactions. 
We have shown that the phase transition from an excitonic insulator to a band insulator occurs by increasing $U^{ab}$ and the region of the excitonic insulator is extended for larger $U^{aa}$.
We have also calculated the exciton pair amplitude both in $\bm{k}$-space and in real space and proposed that the BCS-BEC crossover occurs around $U^{ab}/t\sim 4.8$ for $U^{aa}/t=8.0$.

The extension of the VMC study for more realistic models is straightforward. 
We have checked that the introduction of a long-range Coulomb interaction greatly extends the region of the excitonic insulator phase in the phase diagram~\cite{Watanabe2}.
This is because the long-range Coulomb interaction enhances the long-range exciton pairs and leads to a more stable excitonic insulator phase.
In such a case, the charge Jastrow factor $P_{\mathrm{J_c}}$ becomes more important because the screening effect of long-range Coulomb interaction, which is an essential element for the exciton condensation,
will be properly included through $P_{\mathrm{J_c}}$.
On the other hand, the exciton condensation is greatly suppressed when the perfect nesting condition is broken~\cite{Watanabe2}.
Understanding the effect of the long-range Coulomb interaction and imperfect nesting on the stability of the exciton condensation and other competing phases such as a superconductivity is important to clarify
the detailed properties of TiSe$_2$ and other low carrier density systems.


\section*{References}
\begin{thebibliography}{99}
\bibitem{Mott} Mott N F 1961 {\it Phil. Mag.} {\bf 6} 287
\bibitem{Jerome} J\'{e}rome D, Rice T M and Kohn W 1967 {\it Phys. Rev.} {\bf 158} 462
\bibitem{Halperin} Halperin B I and Rice T M 1968 {\it Rev. Mod. Phys.} {\bf 40} 755
\bibitem{Monney} Monney C, Monney G, Aebi P and Beck H 2012 {\it New J. Phys.} {\bf 14} 075026
\bibitem{Cazzaniga} Cazzaniga M {\it et al.} 2012 {\it Phys. Rev. B} {\bf 85} 195111
\bibitem{Cercellier} Cercellier H {\it et al.} 2007 {\it Phys. Rev. Lett.} {\bf 99} 146403
\bibitem{Hughes} Hughes H P 1977 {\it J. Phys. C: Solid State Phys.} {\bf 10} L319
\bibitem{vanWezel} van Wezel J, Nahai-Williamson P and Saxena S S 2011 {\it Phys. Rev. B} {\bf 83} 024502
\bibitem{Zenker} Zenker B, Fehske H, Beck H, Monney C and Bishop A R 2013 {\it Phys. Rev. B} {\bf 88} 075138
\bibitem{Morosan} Morosan E {\it et al.} 2006 {\it Nature Phys.} {\bf 2} 544
\bibitem{Kusmartseva} Kusmartseva A F, Sipos B, Berger H, Forr\'{o} L and Tuti\v{s} E 2009 {\it Phys. Rev. Lett.} {\bf 103} 236401
\bibitem{Watanabe1} Watanabe H, Shirakawa T and Yunoki S 2014 {\it Phys. Rev. B} {\bf 89} 165115
\bibitem{Sorella} 
  Sorella S 2001 {\it Phys. Rev. B} {\bf 64} 024512; 
  Yunoki S and Sorella S 2006 {\it Phys. Rev. B} {\bf 74} 014408; 
  Watanabe H, Shirakawa T and Yunoki S 2013 {\it Phys. Rev. Lett.} {\bf 110} 027002
\bibitem{Zocher} Zocher B, Timm C and Brydon P M R 2011 {\it Phys. Rev. B} {\bf 84} 144425
\bibitem{Kaneko} Kaneko T, Seki K and Ohta Y 2012 {\it Phys. Rev. B} {\bf 85} 165135
\bibitem{Watanabe2} Watanabe H, Seki K and Yunoki S {\it in preparation}
\end{thebibliography}
\end{document}